\documentclass[letterpaper,twocolumn]{article}

\usepackage{usenix,epsfig,endnotes}
\usepackage{calc}
\usepackage{lipsum}
\usepackage[titletoc, title]{appendix}
\usepackage{tocloft}
\usepackage{fancyvrb}
\usepackage{comment}
\usepackage[utf8]{inputenc}

\usepackage{algpseudocode}
\usepackage{algorithm}

\usepackage[font={small, sf}, labelfont=bf]{caption}
\usepackage{chngcntr}

\usepackage{amsmath,amssymb,amsfonts}
\usepackage{graphicx}
\usepackage{textcomp}
\usepackage{bmpsize}
\usepackage{lipsum}
\usepackage[colorinlistoftodos]{todonotes}
\usepackage{xspace}
\usepackage{comment}
\usepackage{appendix}
\usepackage{graphicx}

\usepackage{amsmath}
\usepackage[export]{adjustbox}
\usepackage{xr}
\usepackage{fixltx2e}
\usepackage{pgf-umlsd}
\usepackage{caption}
\usepackage{subcaption}
\usepackage{soul}
\usepackage{gensymb}

\usepackage{float}

\usepackage{attachfile}

\usepackage{listings}

\begin{document}
\date{}
\title{\Large \bf``My sex-related data is more sensitive than my financial data and I want the same level of security and privacy": User Risk Perceptions and Protective Actions in Female-oriented Technologies}
\author{\rm Maryam Mehrnezhad, Royal Holloway University of London, UK \\
(maryam.mehrnezhad@rhul.ac.uk) \and
\rm Teresa Almeida, Umeå University, Sweden, and ITI/LARSyS, Portugal\\(teresa.almeida@umu.se)
}

\maketitle

\textbf{Abstract:} 
The digitalization of the reproductive body has engaged myriads of cutting-edge technologies in supporting people to know and tackle their intimate health. Generally understood as \textit{female technologies} (aka female-oriented technologies or `FemTech'), these products and systems collect a wide range of intimate data which are processed, transferred, saved and shared with other parties. In this paper, we explore how the ``data-hungry" nature of this industry and the lack of proper safeguarding mechanisms, standards, and regulations for vulnerable data can lead to complex harms or faint agentic potential.
We adopted mixed methods in exploring users' understanding of the security and privacy (SP) of these technologies. 
Our findings show that while users can speculate the range of harms and risks associated with these technologies, they are not equipped and provided with the technological skills to protect themselves against such risks. We discuss a number of approaches, including participatory threat modelling and SP by design, in the context of this work and conclude that such approaches are critical to protect users in these sensitive systems. 


\section{Introduction}
Emerging technologies such as sensor-enabled and connected devices have already found their way into several aspects of our lives including intimate and reproductive health. 
Examples of such technologies include mobile apps for fertility and period tracking, online forums and social media groups for menopause support, and smart IoT devices for nursing such as a breast pump. 
In such sensitive contexts, the introduction of modern technologies may either advance equity or enable multi-layered harms and risks such as
discrimination, bias in algorithms and datasets, and gender-based harassment and violence.
Such harms are complex and have less clear definitions in the literature and can differentially put users at risk. 

In this paper, we focus on digital technologies and the intimate body, specifically digital technologies for reproductive health. These include the latest developments in and categories of Female-oriented technologies (FemTech)\footnote{The majority of the FemTech industry talks about its users as `women'. Nonetheless, queer and trans-venture-backed companies are gathering momentum and contributing to counter inequality and provide safe and reliable health and care to members of diverse and marginalized groups.}. FemTech, a category referring to software, products, and services that advance women’s health but that also stresses the commercial value of female technologies \cite{rosas}, comprises a wealth of female technologies and covers a wide range of intimate bodily topics, such as reproduction. Indeed, the digitalisation of the reproductive body, which has contributed deeply to widening knowledge around intimate health care issues that were neglected or stigmatized, such as menstrual care or sexual health, can also put users at risk and harm, for example, by tracking abortion or infertility \cite{mehr}. The use of digital reproductive health technologies can lead to their users finding themselves threatened and in real danger, such as in violation of (women’s) human rights. For instance, many of such consumer-facing technologies fail to abide by the General Data Protection Regulation (GDPR) and Health Insurance Portability and Accountability Act (HIPAA) in that they continue to share valuable personal information with commercial entities in ways in which users may not be fully aware \cite{lupton1}, for instance, tracking of pregnancy leading to pregnancy-based redundancy \cite{Maternity}, where pregnancy-related discrimination still fails to guarantee effective job protections i.e. not being terminated or demoted due to taking parental/maternity leave. Previous research shows that there is a lack of clarity in the law in relation to this extremely sensitive data, and it is not clear who has the rights to these data, or who could legally access them \cite{shipp2020private}. 

The significance of SP in intimate tech-mediated interactions, in particular, that which traverse sexual and reproductive health care that are a subject of and subject to the law, is critical to safeguard the lives of many. While laws may vary depending on e.g., country or systems of governance, there is a notable concern that insufficient regulatory attention has been devoted to this growing market, and that ambiguity in how the industry is regulated leaves its users at risk of relying on technologies of as-yet unproven worth \cite{McMillan}. The combination of the "data-hungry" nature of this industry and the lack of proper safeguarding mechanisms, standards, and regulations that account for these vulnerable data lead to {new and complex harm} where the user's security, privacy and safety can be at risk. Such harms result from a complex interaction of societal processes driven by diverse stakeholders. 
Consequently, user agency is increasingly fainting in such complex systems where the user does not understand many parts of such an ecosystem. 
That is, users of these technologies are not currently empowered (e.g., through what are readily available technologies and tools) to gain a sense of understanding, control, ownership, security, and trust in using such technologies without risk and fear. 

In this paper, we aim to understand user's SP perceptions, actions, and expectations in FemTech. Our research questions include: \\
\textbf{RQ1}: What is a user's general understanding of these systems i.e., how do the users describe FemTech's functionalities, pros and cons, and their lived experiences of them?\\
\textbf{RQ2}: How do the users perceive the associated and potential harms and risks i.e., what is the user perception of the collection and sharing of data in FemTech?\\
\textbf{RQ3}: What protective actions do users adopt and what are their desired SP features in these technologies?

We conducted mixed-methods user studies to investigate user perspectives of how such intimate data flows in the FemTech ecosystem, with a focus on what the risks are and what protective actions a user might take to protect themselves against such harms. To do this, we first conducted an online survey which we designed specifically to help us understand the general SP mental models and patterns associated with this range of technologies. We publicised our survey in a way that all potential
users of such technologies, i.e. women and other groups, could take part. 
Then, we reflected on the survey results as widely applicable and explored examples of harms more explicitly by engaging 17 participants in an online story completion activity. This was an activity focused on the feelings and protective actions of the participants when encountering a risky scenario when using these technologies. 
Finally, we conducted six individual interviews with a subset of participants from the previous study to discuss their understanding in more depth.  
We then discuss these issues through a set of approaches for designing new technologies and systems, which can foreground values in the design of personal in intimate data, particularly when designing for and with marginalized and vulnerable populations. 

Our contributions include: (i) designing and deploying mixed-methods user studies to enquire user perception of risks and harms and their SP actions (potential and/or preferable), and (ii) exploring approaches to design that treat SP as core concepts and values to better protect users in these systems.

\section{Background and Related Work}
FemTech is a somewhat controversial term \cite{Joyce,Bannon} understood as a subset of health-related technologies that has specifically been designated to advance and foreground the intimate health care needs of women. As such, while focusing on women's health and well-being to address ``a niche" and historically marginalized people, it has also been understood as an industry prone to exacerbating gender disparity and excluding trans and non-binary communities. Past the troubling terminology, this industry is composed of an ever-growing number of technologies that aim to include a plurality of bodies and genders. These include mobile applications (apps), connected devices and online services covering menstruation, menopause, fertility, pregnancy, nursing, sexual wellness, and reproductive health care, to name but a few categories. The class of technologies is broad, ranging from stand-alone mobile menstruation apps to illness-tracking wearables to IVF (In Vitro Fertilisation) services on the blockchain. FemTech Analytics, a strategic analytics agency focused specifically on this sector, predicts an industry growth over \$75-billion by 2025 \cite{femtech1}. The sector is booming, and along with the numerous opportunities to both monetize and arguably improve women's and other marginalized groups health, these come with the increased security, privacy, and safety risks associated with the collection of intimate health and medical data, as required by many FemTech apps and devices\footnote{While FemTech may (need to) expand beyond reproductive health, in this paper we explore these technologies as they focus on the reproductive space (e.g., fertility tracking, nursing, and menopause).}.

In \cite{Mehr2022}, we suggest that FemTech privacy should be looked at via different lenses. These include the cases where someone has user personal data but the user does not --inverse privacy \cite{erickson2022you}, when peer pressure causes people to disclose information to avoid the negative inferences of staying silent --unraveling privacy \cite{peppet2011unraveling}, when the privacy of others (e.g., child, partner, family, friend) also matters  --collective privacy \cite{almeida2022}, and when systems should also focus on the intersectional qualities of individuals and communities --differential vulnerabilities \cite{mehr}. In our previous work, we have identified multiple threat actors for these technologies \cite{Mehr2022}. These interested parties include, but are not limited to: (Ex-)Partner and Family \cite{WHO,stevens2021cyber,chan2021hidden}, Employer and Colleagues \cite{erickson2022you, brown2021Femtech, Maternity, kerry,brown2020supercharged, Drew}, Insurance Firms \cite{crossley2005discrimination,rosenbaum2009insurance, scatterday2022no}, Cyber-criminals \cite{mehr, Carissa, rosas}, Advertising Companies \cite{collective, International,shipp2020private, AntiAbortion, marketplace}, Political and Religious Organisations \cite {Jessica, pennycook2021shifting}, Governments \cite{AntiAbortion, Jerry, minor, Shoichet, IranUN}, and Medical and Research Companies \cite{Carissa, powles2017google}. Such threat actors may exploit these systems in various ways by performing attacks at different points of the ecosystem e.g., human dimensions, hardware vulnerabilities, attacks on datasets, IoT devices, app and website exploits, etc. A number of system studies have been performed on the SP practices of these technologies. Examples include the analysis of the data collection practices of the period tracking app ecosystem and their policies \cite{shipp2020private}, measuring the tracking practices of fertility apps and their compliance to the GDPR \cite{mehr}, and traffic analysis and policy review (with focus on HIPPA \cite{HIPPA1}) of a subset of iOS apps \cite{erickson2022you}. Very limited work has gone into the SP assessment of FemTech IoT devices \cite{valente2019stealing}. The cybersecurity community has yet to properly investigate the data collection of these ecosystems, (lack of) implemented security and privacy-enhancing technologies (PETs), the existing vulnerabilities, and potential security measures to mitigate them. 

While there is a body of research on the SP risks associated with these systems, the human dimensions of these technologies are not explored well.
Previous research shows how the users of such technologies are already marginalised by SP research, products, and tools. This includes e.g., not having the actual end user in the life-cycle of the development of such systems enabled by traditional threat modelling, and disparity in education and SP knowledge \cite{coopamootoo2022feel,mehrnezhad2022can}. Intersectionality and the personal characteristics of the end users of such products have an impact on these systems, too, which is barely discussed in the literature. 
Measuring these risks comprehensively via system studies and user studies is not straightforward and requires mixed methods. 

The SP of FemTech products can be investigated by looking into IoT hardware, product websites, mobile apps, cloud detests, etc. However, the security and privacy of the user and data in FemTech are more complex than in some other contexts. Furthermore, due to the intimate nature of the data involved in these technologies, social and cultural aspects need to be taken into account too. A combination of multi-layered systems and sensitive data and applications creates a new range of complex harms and risks.  
Complex harms include `harms with a less clear definition', which is within the scope of the UK Government’s White Paper on Online Harms\footnote{gov.uk/government/consultations/online$-$harms$-$white$-$paper}. 
A taxonomy of organizational cyber-harms is suggested in \cite{agrafiotis2018taxonomy} where such harms are grouped within six main categories: Physical/Digital, Economic, Psychological, Reputational, and Social/Societal. 
FemTech risks and harms can fall into all these categories, alone or combined, and beyond. Examples include abuse and loss of life, loss of job, shame and guilt, damaged relationships, and public negative perception, all of which belong to the categories described in \cite{agrafiotis2018taxonomy}. Moreover, plenty of news articles, academic papers, and popular science books (e.g. \cite{Carissa}) have included various examples of complex harms and risks.  
These include influencing elections (e.g., \cite{bergmann2020populism}) and organised social media attacks, or LGBTQI+ support for both (digital and physical) safety \cite{Geeng,DeVito}, women's experiences online outside of Western perspectives \cite{Im,Karusala}, vulnerable populations in the global south\cite{Afnan,Sambasivan,Sambasivan1,Sambasivan2}, and people living with stigmatized conditions \cite{Bussone}.
More examples include cyberstalking, and stigmatised topics such as domestic violence/intimate partner violence \cite{Tseng,slupska2022they}, 
and algorithmic harms \cite{strohmayer2022safety}. 
 
Previous research shows that there are gray areas in the current regulations and policies and their enforcement regarding the protection of such sensitive data \cite{shipp2020private,mehr,erickson2022you,brown2021Femtech,brown2020supercharged}. We specifically highlight the gray areas in the intersection of the general data protection laws and the medical and health ones in \cite{MindGap}.
For instance, it is not clear if and how FemTech data is covered by the ``special category data" within the GDPR framework or how this data can be protected against work discrimination e.g., pregnancy redundancy. 
{Similarly, it is unclear what is classified as ``medical'' category vs. other groups (e.g. health and lifestyle) in other relevant guidelines and laws e.g., the UK Medicines and Healthcare products Regulatory Agency (MHRA) \cite{essen2022health, mehr}}. 
It is indeed not clear which stakeholder or party is responsible for the protection and how each party's practice influence others. Furthermore, the combination of multiple technologies (e.g., mobile apps, IoT devices, cloud storage, AI and ML) in these technologies can also lead to unpredicted and unintended consequences. For instance, in the case of marginalised user groups, collecting data does not necessarily only mean including them in the systems and might put them in more danger of being profiled and fingerprinted online. 
All the above dimensions influence user agency where the end user might feel less and less in control of their data and how they understand, perceive, and practice their SP rights. In this paper, we explore user's SP understanding in FemTech to shed light on the complexity of these systems and their harms from the user's point of view. 

\begin{table*}[t]
\centering
\small
 \caption{Overall design of the study. Each section includes the types of questions from the participants (S = Story, Q = Question)}
 \begin{tabular}{|l|l|l|l|} 
 \hline
\textbf{I-a) Survey (general) } &\textbf{ I-b) Survey (risk-related)} &\textbf{II) Story Completion Method} & \textbf{III) Interviews}\\\hline
 - Description \& usage & - Data type & - S1: Mobile app and ads&- Q1: Data flow\\
 - Pros \& Cons  &- Security incidents&(Period and fertility tracking)& in the ecosystem\\
- Ways of learning&  \& Data access &- S2: Online search and ads/& - Q2: Data access\\
- Incidents &- Protective actions (general, FemTech)&misinformation (Menopause care)&- Q3: Potential\\
 &  - Desired security/privacy features& - S3: IoT/app and other user access&Safeguarding \\
 &&(Nursing)&features\\
 \hline
 \end{tabular}
 \label{Overal1}
\end{table*}

\section{Methodology}
We conducted a user study using mixed-methods (online survey, storytelling, and individual interviews) and each method contributes to answer parts of our RQs.  
Table \ref{Overal1} shows the design of our study. We recruited our participants in the UK. 
According to FemTech Analytics\footnote{femtech.health/femtech-market-overview}, the total amount of FemTech funding was \$19.7 billion in December 2022 (+35\% YoY). The UK has the biggest number of FemTech companies in Europe,  
and comes second globally only after the US (UK: 10.5\%, USA: 43.4\% of the whole market). 
In addition, one of the authors is based in the UK and had access to a few platforms (e.g., UK’s National Innovation Centre for Ageing (NICA) and UK-based FemTech organisation Women of Wearables (WoW)) to disseminate the user study invitations. 
We conducted our study between August 2021 and July 2022. 
This survey and subsequent two-part studies had ethical approval granted by the Ethics committee at the authors' Institution. 
We assigned a number to each participant and performed our analysis via that number in order to anonymize our participants. We report our findings in the paper in an anonymised way. 
As a thank you, we gave a £30 Amazon voucher to the participants who completed all three studies. This was communicated with the participants before taking part in the SCM activity and the interviews.
Participation in this study was completely voluntary and our participants could drop out at any stage, as they were informed about in the invitation email/message, consent forms, and later communications. We highlight that recruiting participants for this work was a challenge and it took us several months to complete these studies, mainly due to the lack of engagement from a diverse set of participants. However, the repetitive themes emerging from the three studies reassure the confidence in our results.

\begin{table*}[t]
\centering
\small
  \caption{Demography of our participants}
 \begin{tabular}{| l| l|} 
 \hline
\multicolumn{2}{|c|}{\textbf{Survey (102 Participants)}} \\\hline
Age and Gender& 18-79 years old, 87 Female, 14 Male, 1 Non-Binary\\\hline
Job & (PhD/PG/UG) student (18), university staff (25), private sector job (22),\\
& self-employed/free-lance (16), to not working (11) and retired (10)\\\hline
\multicolumn{2}{|c|}{\textbf{Story Completion Method (SCM) (17 Participants)}} \\\hline
Age and Gender & 22-71 years old,  13 Female, 4 Male\\\hline
Ethnicity& 
White British (9), White European (1), Black African (3), Middle Eastern Asian (4)\\\hline
Job & (PhD/PG/UG) student (5), university staff (3), private sector job (3), \\
&self-employed/free-lance (1),  not working (2) and retired (3)\\
 \hline
 \multicolumn{2}{|c|}{\textbf{Individual Interviews (6 Participants)}} \\\hline
Age and Gender & 22, 28, 30, 32, 36, 71 years old, 5 Female, 1 Male\\\hline
Ethnicity & White British (3), Black African (2), Middle Eastern Asian (1)\\\hline
Status& Single (2), married/ with partner (4), with children (2), trying to conceive (1), postmenopausal (1)\\\hline
Job & PhD student (3), university staff (1), not working (1) and retired (1)\\\hline
 \end{tabular}
 \label{Demo}
\end{table*}

\subsection{Method 1: Survey}
We designed an online survey (Appendix A) to gain a general understanding of user practices in FemTech and their SP concerns and protective actions.

\paragraph{Design Approach}
Our survey had two sections. In the first section (\textit{General Questions}), after a brief introduction, we included questions about the general usage of FemTech and the participants' experiences in their own words. We then asked them about the advantages and disadvantages and how they have learned about those technologies. 
In the second part (\textit{Security and Privacy}), we asked questions about the SP of FemTech. We asked what type of data (e.g., name, age, contact info, lifestyle, sex activities, medical records, etc.) they input into these systems and who they think has access to this data. We then asked what forms of protective actions they take. Finally, we asked our participants, in their opinion, what sorts of SP features should exist in these technologies. 

\paragraph{Participants} We distributed our survey via the newsletter of VOICE organisation\footnote{voice-global.org/}, which is a part of UK’s National Innovation Centre for Ageing\
NICA\footnote{uknica.co.uk/} working with researchers, industry, and end users across the UK and beyond. 
We also distributed our survey via WOW\footnote{womenofwearables.com/} which is a UK-based leading global organisation active in FemTech. We also recruited survey participants via social media and mailing lists.
102 participants completed our survey. At the end of the survey, we asked our participants if they were willing to take part in the further parts of our studies by providing us with their email addresses. This section was voluntary, and 70 participants left their emails for further communication.

\paragraph{Data Collection and Analysis} Our survey was designed in Google Forms. Our method of processing the collected data was a mix of quantitative and qualitative analysis. The results for some of the questions are presented by stacked bar figures. For our free-text style questions, we conducted a thematic analysis \cite{braun}; taking an inductive approach and allowing the data to determine our themes. 
We facilitated a conventional line-by-line coding \cite{line} of all the responses. Both authors contributed to the analysis by coding and extracting the key themes independently. For more complex and lengthy comments, we assigned multiple themes to them. The researchers discussed these themes to agree on potential inconsistencies and also chose the user comments that represent such themes for inclusion in the paper.  

\subsection{Method 2: Story Completion}
\label{Method2}
In this part of our study, we used the story completion method (SCM). SCM is a method that allows for creative enquiry and imaginative engagement with one’s own experience as well as trying out someone else’s shoes and can encourage people to consider the implications of social problems in ways that do not involve direct questioning \cite{Ash}. As a research method, story completion is a generative technique for eliciting narrative data that does not involve personal or direct questioning \cite{clarke2019editorial}.
This qualitative research method has also seen a growing interest when researching sensitive topics, in particular areas of research in which collecting data could be more difficult face-to-face, or ethically challenging. For example, in \cite{Wood}, the authors explore VR and technosexuality and suggest SCM as a critical tool to enquire and problematize the status of technology, or \cite{Troiano} whose stories explore humans encountering and interacting with sex robots, the stories solicited both ways from the human as well the sex robot’s perspective. In \cite{moniz2023intimate}, we highlighted SCM as a speculative approach that is valuable and important when the aim includes asking questions around the sociocultural and ethical implications of intimate and sensitive contexts and technologies.

\paragraph{Design Approach} 
This activity involved the participants in speculating the user's thoughts and actions within a variety of FemTech-related and potentially risky-scenarios by engaging in the creative process of completing a story. We presented participants with a set of three story prompts (`stems’) for them to complete. These story stems involved scenarios using digital technologies, to produce written narratives about situations faced by fictional protagonists. 
Each of the scenarios was created based on the SP concerns and protective actions reported by our participants in the survey. As we will discuss at length in the result section, our survey participants identified data sharing as their biggest concern, regardless of the type of the product and its applications. We reflected such concerns in the design of our stories. These stories were carefully designed across the range of FemTech products (mobile apps, online websites, IoT devices) and applications therein (fertility tracking, menopause aid, nursing), and potential risks (user profiling and tracking, misinformation, and violating others' privacy). Our story prompts include (the format of this activity is included in Appendix B):  

\begin{itemize}
    \item \textbf{Story 1 (mobile apps and ads)}: Alex downloads a free period tracking app to keep an eye on their fertility window. The next day Alex receives multiple email ads about baby’s clothing and toys. What does Alex think? What happens next?

    \item \textbf{Story 2 (Internet search and ads)}: Morgan goes online and browses the internet, searching for common symptoms of menopause. The next time Morgan logs on to Instagram and Facebook, both feeds are full of ads for boosting sex drive and weight loss. What does Morgan think? What happens next?

    \item \textbf{Story 3 (IoT device/app and user access)}: Ashley has been using a ‘smart’ breast pump that connects to an app and is shared within the family. This is a breast pump that can track breastfeeding and give tips to support parents in meeting  their goals. In the meantime, Ashley’s partner has received an alert regarding the baby’s feeding schedule and weight and feels upset that it is not according to their plans. What does Ashley think? What happens next?
\end{itemize}

These three story prompts adopted gender neutrality in the naming of the three fictional protagonists and their respective hypothetical scenarios. We designed our stories in a general way to allow the participants to interpret the context freely i.e. they can think the actor in the story is using such services (app, device, online search) for themselves or for someone else. While this general approach was intentional, we acknowledge that the results might lead to diverse patterns. However, our results show that some general patterns emerge from these stories.  


\paragraph{Participants} Following the initial survey, we invited the 70 participants who declared an interest in our follow-up studies to take part in this activity. We sent the participants an email invitation and a link to this activity. 17 participants completed the activity in an anonymous way. Participants were asked to spend a maximum of 15 minutes writing on each scenario and there was no maximum or minimum amount of words requested.

\paragraph{Data Collection and Analysis} 
We created a Google Form and shared it with our participants. This form included the three story prompts designed for the participants to first, reflect on what someone else might `think' and second, the potential action that someone might take. The minimum amount of words entered was four (``don't like, uninstall it") and the maximum was 269. Similar to the previous section, we used a process of thematic analysis to extract themes as we report in the results.  
This method provided our participants with an excellent opportunity to think about the end user of the same products that they use and/or don't use, their feelings, risks, and potential protective actions. 

\subsection{Method 3: Individual Interviews}
Our final method included one-to-one online interviews with our participants and asking them to reply to the questions by drawing their answers as well as describing their sketches. 

\paragraph{Design Approach} The goal of this activity was to reflect on the user's response to the previous activities and have an in-depth discussion about the FemTech ecosystem. More specifically, we wanted to observe the understanding of the user about the data flow and sharing and their understanding of the big picture. This was done via one-to-one interviews by asking our participants to draw some parts of their answers and thoughts about these questions. Drawing, sketching, and mapping are commonly used to facilitate interviews, and inspire deeper and more thorough conversations \cite{kaye2014money,ryan2009device}. Such methods have been used in SP research too e.g.,
in \cite{vertesi2016data} the risks and harms to individuals are studied in the context of accumulating the digital traces of users across online services and platforms. The interview questions were supplemented by asking the participants to hand-draw sketches of
their digital ecosystem on paper.
In addition, in \cite{nicol2022revealing}, in order to facilitate the interviews on the management of individuals' personal
data, the participants were invited to draw a map of their data and complete this over the course of the interview.

Following the same methods of accompanying our interviews with hand-made drawings, after a brief introduction, we asked three concrete questions from the participants as below: 

\begin{itemize}
    \item \textbf{Question 1 (Data flow)}: In any form of FemTech solution (low tech (e.g., off-the-shelf pregnancy test), shopping website, GP system, mobile app, IoT device), how do you think your data might be transferred beyond your own device? What are the connections in this ecosystem? Can you please doodle something for us (i.e. please draw a general outline of this ecosystem)? Where do you think this data will end up (any possible place)? 
   
    \item \textbf{Question 2 (Data access)}: In such an ecosystem, who do you think might have access to your data (e.g., yourself, partner, family/friends, social media, employer, colleagues, government, GP/NHS, council/police, product company, advertising companies, researchers, general public, etc.)? Please identify them in your sketch and complete it if necessary.
    
    \item \textbf{Question 3 (Safeguarding features)}: In this ecosystem, how do you think user data and their privacy and security can be protected? Please give us three concrete ideas and highlight them in the sketch.
\end{itemize}

\paragraph{Participants} We invited the participants who completed the SCM activity to volunteer for a one-to-one interview with the authors of the paper by sending them an email. Six participants agreed to take part, and the interviews took place within two weeks of the SCM activity. While asking each question, we explained to our participants that they were free to think out loud while completing their drawings. We clarified that this doodling exercise was not
intended to produce judgments of expertise. We explained that this was for remembering the flow of the conversation by referring to specific points of their drawings. 
All interviews took place online and we used the video conferencing tool Zoom to conduct these virtual meetings. Interviews lasted between 30 and 60 minutes. 

\paragraph{Data Collection and Analysis} 
We recorded the online interviews with the participants using the video conferencing tool Zoom. The interviews were transcribed by one of the authors, they were anonymised by assigning a number to each participant and no personally identifiable information was kept on file. Similar to the previous sections, we conducted an inductive thematic analysis to extract the emerging themes by assigning preliminary codes and searching for patterns across the interviews.
During the interview, we asked each participant to draw a sketch on paper and send a scan or image of it to us via email. Following this activity, we ended up with six sketches, in addition to the user comments on their drawings, reflecting the user's understanding of these systems, the stakeholders, data flow and access, and suggested mechanisms to reduce risks.  
We analysed these drawings (components, data flow/access, and suggested SP features) according to a general IoT ecosystem and its multiple components including IoT devices/apps/websites, cloud storage, third party connections, etc. 
 
\subsection{Limitations} 
We composed our research around the general term ``FemTech" which is applied to a wide range of technologies in our studies while explaining that this is a term used by the industry sector and these systems are not limited to women.  
We acknowledge that the range of FemTech is wide. However, our previous research shows that they all collect a similar range of sensitive user data \cite{almeida2022,Mehr2022,MindGap}. 
Our stories and interview questions were designed in a general way for the purpose of this study to understand the general patterns associated with user risk perceptions and actions in FemTech.

\begin{table*}[t]
\centering
\footnotesize
 \caption{Users views on the advantages and disadvantages (Survey)}
 \begin{tabular}{|l|l|l|l||l|l|l|l|} 
 \hline
\textbf{Advantages} & User \%&\textbf{Advantages} & User \%&\textbf{Disadvantages} & User \%&\textbf{Disadvantages} & User \%\\\hline
Easy to use & 72 &None-intrusive to body & 43&Share user information & 65&Biased & 26\\
Increase knowledge & 60 & Cost-effective& 42&Not accessible to all & 29 &Expensive & 10\\
Personalised features & 56& Accurate& 18 & Not easy to use &28 &Others& 5\\
Easy access to all& 48 &Others& 5& Not accurate& 28&& \\
 \hline
 \end{tabular}
 \label{advsdis}
\end{table*}
We plan to use each of these methods with more depth and detail in the future to uncover more nuanced results. We report more detailed demographic information for interviews only since these very personal questions (e.g, relationship and parenthood status) were only extracted during one-to-one interviews in an organic way rather than direct questions. 
We conducted our studies in the UK, which has a diverse population and is a country of multiple identities in Europe. We acknowledge that further studies are required across different user groups and we leave that as future work.

\section{Findings} 
Generally, participants in the online survey study described FemTech as a series of technologies supporting female and women's health including ``physical, mental and sexual health". Examples (free-text questions) included \textit{period}, \textit{fertility}, \textit{ovulation} and \textit{pregnancy tracking}. Our participants chose a number of ways in which they have learned about FemTech including personal research (46\%), partner/family/friend (34\%), online ad (24\%), colleague/work (10\%), school/university (8\%), health centre/clinic or NHS (7\%), TV/Radio (4\%), and traditional ads (1\%). 

When we asked about the advantages of these technologies (via itemized questions as well as free-text boxes for extra items, demonstrated in Table \ref{advsdis})
, 72\% believed that ``They are easy to use", 60\% chose ``They increase my knowledge about my/partner's body", 56\% said ``They can be personalised", 48\% chose ``Everyone can have access to them", 43\% believed "They are non-intrusive", 42\% said ``They are cost-effective", 18\% chose ``They are accurate". Those who chose the ``Others" category mentioned other advantages such as ``\textit{[FemTech] help partners and family for planning various things such as trips, and parties}" and ``\textit{[They are] relatively anonymous}".

\subsection{Perceived Harms and Risks}
\subsubsection{General Perception} Via different parts of our studies, we asked our participants questions relating to their understanding of the risks and harms associated with FemTech. For instance, we asked about the advantages and disadvantages of these technologies as reported in Table \ref{advsdis}, 
as well as their lived experiences regarding any incidents when using these technologies.

Our participants chose a number of issues that they believed to be the disadvantages of such technologies. 65\% of the participants believed that ``They share user information with third parties or sell to them", 29\% chose ``They are not accessible to all user groups", followed by 28\% choosing ``Using such technologies is not easy for certain users", 28\% choosing ``They are not accurate" and another 26\% believing that ``They only work for certain user groups and have bias", and 10\% said that ``They are expensive". Among those who chose ``Others" (5\%), a few said that Governments or other entities might ``track women" via these technologies, and others said these technologies might change what is considered as ``\textit{normal}". 
 
On the experiences regarding any unpleasant incidents, among the 12 replies, three said the inaccuracy of FemTech has led them to lose their fertile days and/or one has experienced an unwanted pregnancy (``that did not match my life plans timeline") and one said he has heard of an unwanted pregnancy of a friend due to ``trusting the calculation of the none
-fertile days via an app."
For example, one participant noted ``\textit{My wife lost a few months when she wanted to get pregnant since the app was not accurate about the fertile days}". Another said: ``\textit{I have heard of women being overdue with their period and the algorithms showing them advertising for baby stuff before they even managed to tell their partners or anybody else. Not ok.}"
Five users said their privacy and security have been at risk e.g., they have noticed that they have been tracked via seeing personalised ads without their consent. Other examples given by users included not having a backup of their data and leaking their data. Finally, two participants said that these technologies set the bar too high and cause stress for the users e.g., regarding what is a ``\textit{normal period}" and what ``\textit{milk volume}" is enough for a baby. For instance, one participant said: ``\textit{They can be misleading and often put unwanted pressure on a person that their body should be behaving in a certain manner. Without these apps, there was less awareness and perhaps less stress and anxiety around ``what is normal”}".

As can be observed in these responses, some of the immediate risks were of concern to a greater percentage of our participants. For instance, most of our participants (65\%) could infer that sharing user data can be an issue in these technologies. This concern was also reflected in users' lived experiences (extracted from users' comments). This particular concern informed the design of the stories of study 2.
In addition, around 30\% of the participants believed that these technologies are not accurate and one quarter believe that they can be biased. These can create serious harm to the user e.g., unplanned pregnancy, experiencing anxiety in conceiving, wrong mentality about body norms, etc. 

\subsubsection{Data and Access}
We also directly asked our participants about the types of data collected in these technologies as well as who might access this data. We listed a range of FemTech data and allowed the users to choose them as well as add items which were not listed. This list had a few items informed by our previous studies (\cite{almeida2022,Mehr2022,MindGap}) including: period-related info (chosen by 60\% of the participants), user info (name, age, gender) (50\%), contact info (40\%), emotional and physical symptoms and lifestyle information (each 35\%), sexual activities (25\%), medical information, pregnancy information and information about reproductive organs (each 10\%), users' social media, baby, photos (each 5\%). We presented this list to the users to remind them of the wide range of user data that can be collected by these technologies. Our previous work \cite{almeida2022,Mehr2022,MindGap} shows that these technologies collect all the categories intensely. We observed that such data collection is more than what our participants reported. This is the case for most data types (self-entered, automatic, sensors) across the majority of FemTech categories. 
\begin{figure}[t]
\includegraphics[scale=.32]{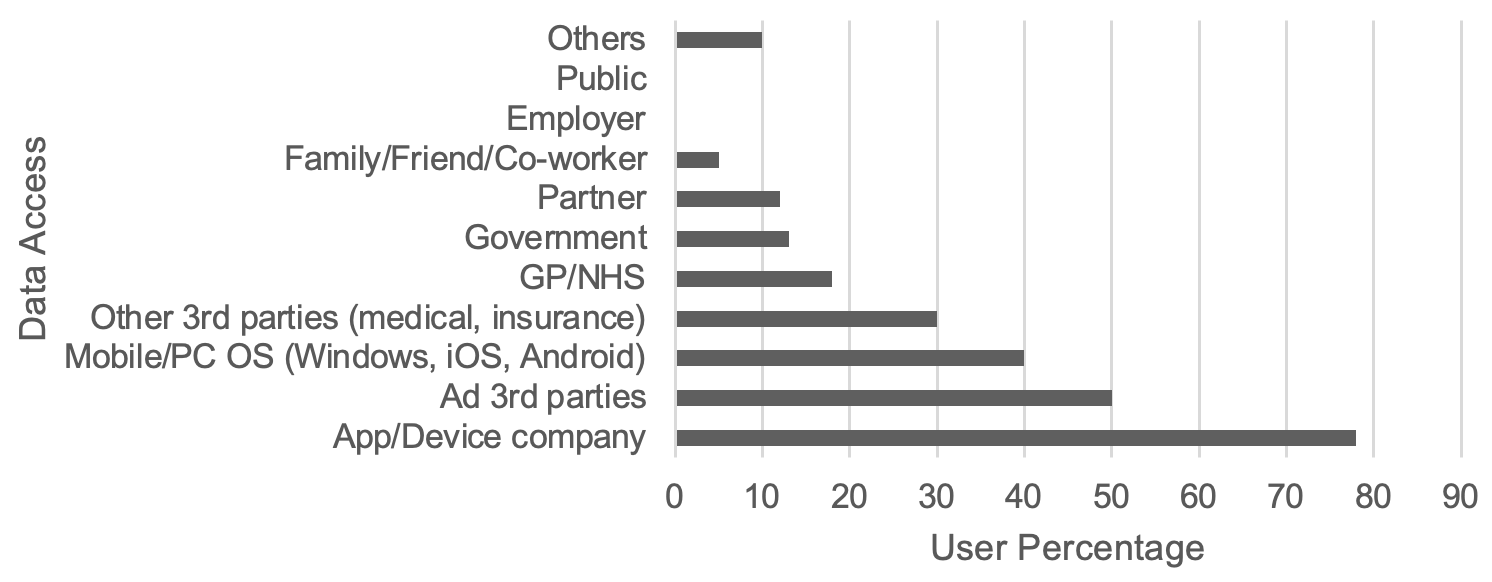}
\caption{Users' views on data access in the Survey. The y axis represents the entities who might have access to FemTech data.}
\label{Access}
\end{figure}
We asked our participants who they think has access to this data. We provided a list, partially informed by previous research (ibid), and allowed them to add new items via a text box. As reported in Fig. \ref{Access}, after the app/device company (78\%), the ads companies category was the most selected one (50\%). 40\% of the participants chose the OS of the hosting device, and 30\% picked third parties such as medical clinics and insurance companies. Over 10\% believed that GP/NHS, government, and/or partner can have access to this data. Those who chose others mentioned different entities e.g., ``\textit{whom the company decides}" (five participants) and ``\textit{anti-abortion groups}" (two participants). In our previous work, we have indeed highlighted that all these entities are interested in such data and the data collection and sharing is not always transparent in these technologies \cite{almeida2022,Mehr2022,MindGap}.

We also asked the participants if they had ever experienced any SP incident when using such technologies. Among the 40 participants who provided answers to this question, 10 participants said no, 15 said they are \textit{``not sure}", ``\textit{it is possible}", however, they ``\textit{might not be aware}" of such an incident. For instance, one said: ``\textit{I'm not sure how I would tell. I pretty much expect that somewhere in the fine print I have implicitly agreed to all kinds of pernicious data sharing...}". 15 participants listed specific incidents. Eight of them said they have been bothered by personalised ads, four said they had not given any explicit consent for their sensitive data to be collected and shared by these apps, and three said they had experienced a password leak. For instance, one participant said: ``\textit{As advertisement have been sent to me I am sure third parties access to my data that I had entered.}" Another one said: ``\textit{Yes, I received personalized ads. I also don't see any privacy alert according to GDPR}".

Throughout our user studies, we aimed to gain an understanding of the user perception of the FemTech ecosystem and the data access and flow beyond the user's device. This was reflected in our online survey and further on in our one-to-one interviews. As previously discussed, around half of the participants in the online survey recognised that different stakeholders (e.g., third parties) might have access to FemTech data (Fig. \ref{Access}). In addition, all six participants in the interviews also recognised that such data could be transferred to third parties via the FemTech company and with or without the user's consent. This was reflected in their drawings. Despite that, these six participants had different views of how other interested parties (e.g., NHS, government, insurance companies, etc.) may or may not have access to FemTech data. Fig \ref{Eco} (Appendix C) shows two examples of such drawings. As one can see, these participants anticipate that multiple third parties may have access to this data and re-share it with others too. Through our discussions, we identified a high level of uncertainty among our participants about how this ecosystem works and what risks threaten them. This was also reflected in our online survey where multiple users said that they don't know if and how their data is shared beyond their device. All these six participants also highlighted that they do not know if this data is protected by the relevant law e.g., the GDPR and were hesitant to speculate on all the potential parties and their access to such data. 

\begin{figure}[t]
\includegraphics[scale=.32]{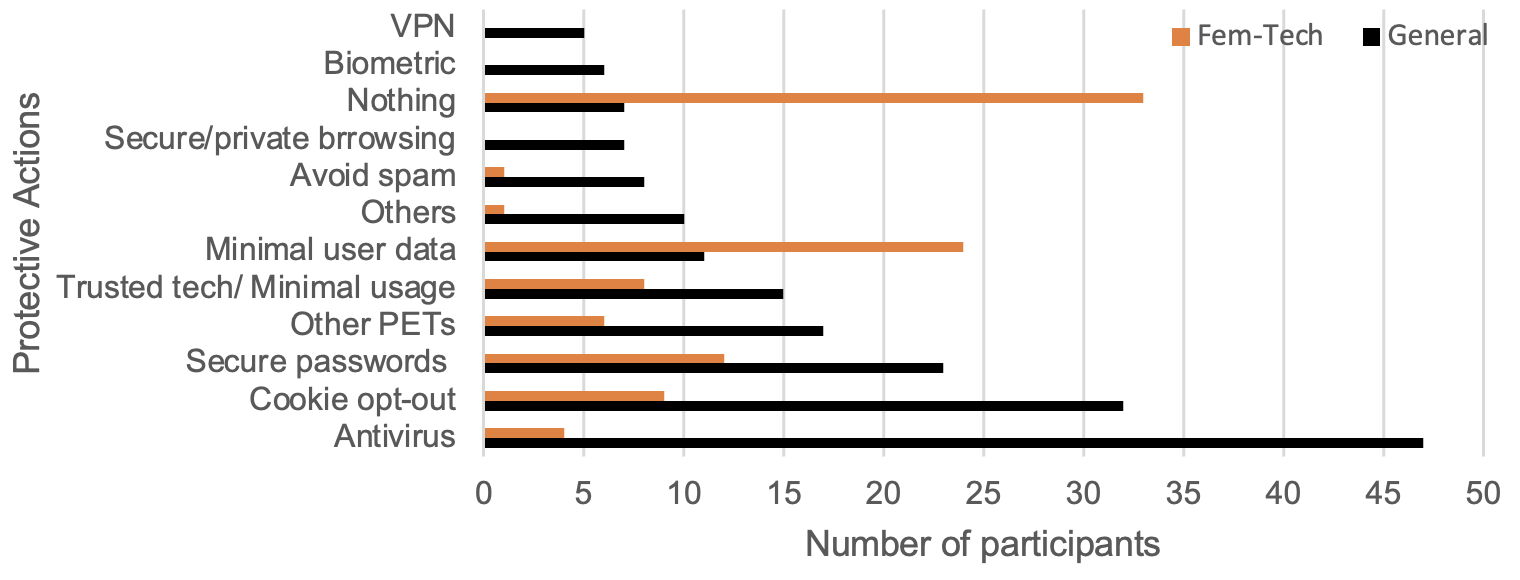}
\caption{How users protect their privacy and security in general vs. FemTech (Survey).} 
\label{Actions}
\end{figure}

\subsection{Reported Protective Actions}
\begin{table*}[t]
\centering
\small
 \caption{Feelings and protective actions in SCM (NA: no feelings were described in the story)}
 \begin{tabular}{|l|l|l|l|} 
 \hline
\textbf{Story}& \textbf{Actor} &\textbf{Feelings (no. of users)} & \textbf{Actions (no. of users)} \\\hline
  1. Mobile app and ads &Alex&Negative (14), Neutral (2), & Uninstall/Switch app (11), Review settings (4), \\
  &&NA (1)&Others (4)\\ \hline
  2. Online search and ads& Morgan&Negative (13), Neutral (1),&
  Review browser/cookies settings (7), Limit/Stop  
  \\ && NA (3)& internet search (6), Click ads (5), Others (3)\\ \hline
  3. IoT device/app and access &Ashley&Negative (15), Neutral (1), &Review settings (8), Stop using (6),  
  \\&&NA (2)&Discuss with partner (6), Seek help (6)\\
 \hline
 \end{tabular}
 \label{Feelings}
\end{table*}
\subsubsection{General Protective Actions} 
We asked our survey participants about the ways that they generally protect their online SP in FemTech in two open-text questions. Fig. \ref{Actions} shows the themes that were visible in the answers provided by our participants. As one can see, 47\% of the participants use Antivirus and more than 30\% opt-out of cookies to protect their general online SP. Other methods include secure passwords (e.g., password managers, changing passwords, different passwords for different accounts) (23\%), and other forms of PETs such as Tor, DuckDuckGo, and reading privacy policies (17\%). 15\% of our participants said that they only use a trusted app/tech and/or only use the technologies that they need in a minimal way. 11\% said they only input the minimal data that will make the app/tech function well. 
10\% of the participants said they use other ways of protecting themselves via e.g., personal research, providing false data, using multiple accounts, managing social media, system updates, etc. 8\% said they make sure they don't fall victim to a spam attack, and 7\% said they use secure browsers and private browsing via e.g., incognito mode. 6\% the participants said they use the biometric features of their mobile devices to protect themselves and 5\% said they used VPN. Only around 7\% said they do nothing to protect their online SP. 

In contrast, 32\% of the participants said they do nothing specifically to protect their FemTech privacy and security. 
Providing minimal user data into these technologies was the most popular method, selected by 24\% of the participants. For example, one participant said: ``\textit{not entering too much info especially medical and sex-related ones since they are super sensitive}". Another participant said: ``\textit{I do not share anything else than my period details. I do not feel safe and doubt these applications.}" Other methods included antivirus, cookie opt-out, secure passwords, other PETs, and using trusted technologies. For instance, one participant said: ``\textit{Use only limited ones. Not third party downloads except NHS app, though that's likely to have the same data issues, but have to accept this. Opted out of data sharing from NHS via govt website.}" 
As it can be seen Fig. \ref{Actions}, most protective actions are general practices for the users rather than specific to FemTech. The only exemption is providing ``minimal user data". This protection method is interesting since it implies that these participants realise the intimate nature of FemTech data and do not see any benefit in providing such data to these apps. For instance, one participant said: ``I only provide minimum data to make the app work with a fake name and nothing more."

\subsubsection{Potentially Risky Scenarios} 
We also analysed the data collected via our SCM. 17 participants completed this activity and reported their thoughts of potential security or privacy invasions as described in each story (Section \ref{Method2}). The results are reported in Table \ref{Feelings}.
While we asked our participants about what would the user in each story think and what happens next, they mainly reported on feelings and protective actions. The majority of the participants believed that the users of these stories would feel negative about the experience. This was reflected by the use of terms such as \textit{``angry", ``annoyed", `upset", ``incandescent", ``spied on", ``not in-control", ``not empowered", ``worried", ``concerned", ``tracked", ``bothered", ``shocked", ``irritated"}, etc. For instance, one of the participants wrote: ``\textit{Alex sees the emails and is concerned about their phone monitoring what they do. Although this isn’t the first time they’ve received ads about things they’ve recently searched or downloaded, the more personal and sensitive aspect of this application has them worried. This application captures information about their health, Alex views this information as sensitive and personal and doesn’t feel comfortable with any aspect of this being shared. ...}".
In terms of the actions that the users of these stories might take, our participants came up with multiple ideas. In the case of the use of the period tracking app and baby clothing/toys ads, the majority of the participants suggested that Alex would uninstall the app or switch to a new app with better SP features. Four participants said that Alex would review the app settings or their Google privacy settings and make modifications. 
In the case of the second story, Morgan searching the internet for menopause symptoms, seven participants said that Morgan should review their browser settings and/or opt-out of cookie notices for a more private search experience. Six participants said that this experience would make Morgan more cautious and they might limit or stop their use of internet searches. Three participants mentioned that Morgan is likely from an older generation and is less confident and equipped to take any protective actions. Alex and Morgan were referred to mainly as she, but sometimes as he in these stories and one time in a non-gendered way.
In the case of Ashley and the shared use of a smart breast pump app, eight participants envisioned that Ashley would review the settings and modify the access of their partner. Six said that Ashley would stop using the device and app altogether, another six said that they would have a discussion with their partner about app privacy setting and/or their goals for feeding the baby, and another six said that they would seek help from GP, family, friend, etc. to support their breastfeeding goals. As one can see, the majority of these protective actions are related to protecting the privacy of the users. Ashley was mainly assumed to be female in the stories, one time male and one time in a non-gendered way.

Previous research (e.g., \cite{coopamootoo2022feel}) suggests that security and privacy-protective actions are closely associated with (and predicted by) particular negative feelings in the context of online tracking on websites. Our results are in line with those findings where the majority of the SCM's participants describe negative feelings for the story actors and then suggest concrete protective actions. While we did not directly ask about the feelings of our survey participants, 65\% of the participants chose ``share user information" as the main disadvantage of such technologies. When we analysed our data, we found out that those participants also chose more items for protective actions (both in general and for FemTech, Fig. \ref{Actions}). While further studies and more participants are needed to confirm these results, we believe that the participant's protective actions in these intimate digital health technologies are also associated with their negative feelings. 

\begin{table*}[t]
\centering
\small
 \caption{Desired security and privacy (SP) features (Survey)}
 \begin{tabular}{|l|l|l|l|} 
 \hline
 \textbf{SP Feature} & \textbf{User \%} & \textbf{SP Feature} & \textbf{User \%}\\\hline
Security mechanisms 	&30&Not share/sell, local storage	&16\\
(password, encryption, biometrics, anonymisation, etc.)&&Minimal data collection	&7\\
Transparency, enforcement guarantee	&28&Trusted entities	&3\\
User consent, user control	&24&By design (security, privacy, expert, women)	&3\\
No idea, not important, not possible	&20&Others (cloud backup, education)	&3\\
 \hline
 \end{tabular}
 \label{Features}
\end{table*}

\subsection{Expected Security and Privacy Features}
We concluded our survey by asking what types of SP features the participants would like to see in these technologies. This was an open-text question. We analysed the responses and extracted multiple desired features (Table \ref{Features}).  
30\% of our participants suggested various forms of security mechanisms including multi-factor authentication, encryption, biometrics, and anonymisation. For instance, one participant said: ``\textit{User login details should be handled securely, use multi-factor authentication for account set up and potentially for login if the app handles highly sensitive data. An easy to understand privacy policy that is clearly displayed to the user and outlines that user data isn't shared should alleviate user concerns regarding the privacy of their data}". 

28\% participants expected to see more transparency at various stages of data collection, processing, transferring and sharing in FemTech. One participant said: ``\textit{I think they should be transparent about their algorithms and data sharing e.g., open-source code}". Another said: ``\textit{It should be clearer who is viewing your data and how this will be used}". Another one said: ``\textit{[there should be] Full privacy guarantees from companies}". 24\% of the participants said user control and consent should be included in these technologies. Two example comments include: ``\textit{I think the app should clearly explain the use of data to user and provide easy opt-out options}" and ``\textit{A clause to tick if you don’t wish to share information with third parties}". 20\% of the participants said they either don't know or it is not important, or it is not possible to protect user data in such technologies. A participant said: ``\textit{No such thing if someone can hack the White House they can certainly hack these things}". 16\% of the participants said that this data should not be shared or sold. For example, one said: ``\textit{I'm not sure that there can ever be 100\% security/trust, but avoiding selling data to shady third parties is a must.}" 

Other suggested methods include minimal data collection and local storage, anonymisation, development of such products by trusted entities, focusing on the design phase, and others. 
For example, one said: ``\textit{Not requiring irrelevant personal information while signing up to the application. For example, there's no point for your home address information to be required while signing up for a period tracking application.}" Another one said: ``\textit{Everything should be backed up on the cloud anonymous and only the user should access to it}". 
Another comment said: ``\textit{I would feel a lot more secure if something was made by the NHS or something similar, at least they can be held to account. Generally, I want more specific guarantees that data isn't being sold and/or what the app publisher does with the data}". Some other participants said that ``\textit{security researchers}", ``\textit{other experts}", and ``\textit{women}" should be involved in the design processes. 

Similarly, in the SCM activity and our interviews, the participants recognised the level of sensitivity of the data and the required safeguarding mechanisms and protective actions. 
In both the survey and SCM studies, our participants suggested concrete and effective SP methods to be embedded in these technologies.
For instance, in our interviews, the participants believed that more safeguarding mechanisms should be in place to protect this data. The same SP features presented in Table \ref{Features} came up in the interviews e.g., the use of strong passwords, user consent and transparency in the policies, and not sharing the data with third parties. When asked for concrete security ideas for SP protection, two of our participants suggested similar mechanisms adopted by ``\textit{banking systems}" e.g., second-factor authentication and biometrics since they believed that this data is ``\textit{as sensitive}" or ``\textit{even more sensitive}" than financial records. 
{For example, one of our participants said: ``My sex-related data is definitely more sensitive than my financial data. It is very personal and should be protected with the same SP features of banking systems or even more".}
These suggested SP features are particularly interesting since they are distant from those that users are using in practice. As reported in Fig. \ref{Actions}, our participants do not use any significant technology-based protective method for FemTech. Indeed,
the most popular method is providing minimal user data into these solutions -- a method that is only applied by less than a quarter of our participants. The lack of appropriate SP methods in the current practices and their impact on user perception and practice are clearly recognised across different sections of our studies.

\section{Discussion}
In this section, we discuss the results of our mixed-method study and introduce a number of approaches to consider in the design and development of these systems. Our results show that:

(i) Users recognise that FemTech data is sensitive and they feel negative about the (potential) risks of these technologies.  
We surveyed a plurality of people and used open and creative methods for them to explore capacity, choice, and action in relation to this intimate digital health ecosystem. What emerged shows that there is both a need and an opportunity to improve current practices and lessen risks. An example of an approach for doing that includes accounting for human values throughout the design process. While this approach may introduce more challenges for the developers, a `Value Sensitive Design' (VSD) \cite{Friedman,friedman2013value} inspired approach can contribute to improving user experience in the context of FemTech. 
VSD is an approach that has been successfully used in the design of SP features of IT systems, including security features and privacy-enhancing tools in web browsers \cite{millett2001cookies,xu2012value} and ethical design of AI and ML e.g., in conversational agents \cite{wambsganss2021ethical}. Similar work has gone into VSD and user groups such as older adults \cite{wambsganss2021ethical} and children \cite{xu2008alleviating}. 

(ii) Users are not sure how these systems (e.g., apps, websites, data storage, algorithms, etc.) work, what are the data flow and recipients, and what their rights are in the law. 
We suggest SP by design be regarded as a safeguarding paradigm in the design and development of these products. 
Security by design refers to a series of processes and approaches in the life-cycle of the system and calls for the design and development of software products and capabilities to be foundationally secure. Similarly, privacy by design calls for privacy to be taken into account throughout the whole engineering process of a system.  
The seven foundational principles of privacy by design are (1) Proactive not reactive; Preventative not remedial, (2) Privacy as the default, (3) Privacy embedded into design, (4) Full functionality – positive-sum, not zero-sum, (5) End-to-end security– life-cycle protection, (6) Visibility and transparency, and (7) Respect for user privacy \cite{cavoukian2009privacy}. When the above principles are incorporated into the system's life-cycle, they improve the chances of treating each system case by case; as opposed to developing a general system following general methods e.g., threat modelling tools, where key threats to these sensitive systems might not be identified. 
Some of the above principles (e.g., opting in vs. opting out, privacy embedded in the design, encryption, and transparency) were partially mentioned by our participants as their preferred SP features of FemTech (Table \ref{Features}).
Accordingly, incorporating SP by design in FemTech systems can significantly improve these technologies; helping to consider the `actual' end user rather than the `abstract' user throughout the system life-cycle. However, there are complexities in implementing such SP approaches in many contexts, such as digital health, due to the conflict of interest between the company's monetisation of user data and user privacy. We believe that by regulating the sector better, such SP practices can be better realised.

(iii) Participants suggest reasonable and effective SP methods to be embedded in these technologies. These results emphasise the extent to which our participants, i.e. users, have a somewhat confused understanding of how they can be in control of such systems and data and/or are not equipped with the tools and skills to protect themselves from the potential harms. 
We advocate employing 
participatory threat modelling and incorporating marginalised populations’ experiences in the modelling of the threats of a system \cite{slupska2022they,slupska2021participatory}.  
By exploring the range of harms in FemTech, we highlight how the complexity of these risks can make them difficult to be identified by traditional tools and methods. For instance, the nature and dimensions of such threats are sometimes seen as side concerns \cite{Slupska:2019:1746-451X:83}.  
Additionally, by conducting our user studies, we have concluded that these risks and harms can indeed be recognised by the participants. 
For instance, sharing user information was the top disadvantage of FemTech as reported by 65\% of the survey's participants, a risk that would not be treated as an ``attack” in conventional cybersecurity literature \cite{Slupska:2019:1746-451X:83}. 
The confusion that was expressed by our participants across the three studies of how these systems work (e.g., data access and flow) clearly conflicts with the way these technologies are advertised; i.e. empowering the user. Contrarily, these sensor-enabled and smart devices and apps create a new range of threats and risks that did not exist before. Hence, the classic threat modelling tools are not responding to them. Slupska, Duckworth, and Neff \cite{slupska2021reconfigure} discuss how women, particularly women of colour, and queer people are targets of complex harms disproportionately e.g., via online harassment, stalking, and image-based sexual abuse. Threat models are, however, the result of the worldview of security experts as opposed to the lived experiences of the end users. Accordingly, such threats are often dismissed as a ``privacy concern” outside the scope of real cybersecurity research 
\cite{Slupska:2019:1746-451X:83}.

Several recent papers (e.g., \cite{mcmillan2023rethinking}) have shown and discussed how the gray area in different laws (general data protection, health, and medical systems and devices) and lack of coverage for FemTech data provide opportunities to exploit the user data. The ‘Data Protection by Design and by Default' (DPbD) is defined by the GDPR as a requirement for the service providers to put in place appropriate technical and organisational measures to implement the data protection principles effectively and safeguard individual rights \cite{ICObyDesign}.
The ICO website provides a series of guidelines to help the service providers to practice the DPbD. However, even within the available guidelines (which fail to include FemTech data), developers fail to practice them and require support to help them meet the DPbD requirement imposed by GDPR \cite{crabtree2022privacy}. 
We invite the academic community as well as industrial and regulatory bodies to proactively work on the SP of FemTech to enable the delivery of more privacy-preserving products for the end user.

\section{Conclusion}
The variety of products (websites, IoT devices, mobile apps, etc.) in the FemTech ecosystem and the data interaction with diverse end users and other stakeholders create a range of new and complex harms. This introduces more challenges in establishing user agency in these socio-technical systems. In this paper, we explore these issues via an empirical enquiry composed of three mixed-methods user studies in the UK. The results of our online survey (n=102) highlight that 65\% of our participants recognised that sharing user information is a disadvantage of these technologies. 
The results of our story completion method activity reveal that our participants (n=17) mainly felt negative about such data sharing and its consequences. In addition, the results of our interviews (n=6) provide an indication that the users identified the complex nature of the FemTech data and systems. Throughout all these studies, our participants also expressed uncertainty in understanding these systems and risks as well as their rights regarding the SP protection that they need to get from other stakeholders e.g., lawmakers, service providers, and the government.

We discussed incorporating current methods such as VSD and participatory threat modelling for including marginalised populations’ experiences in the design and implementation of these sensitive systems. By integrating such concepts and methods into the general guidelines and approaches for SP by design, the sector can build the next generation of FemTech solutions to be privacy-preserving and secure. 


\textbf{Acknowledgements}: This work has been supported by the PETRAS National Centre of Excellence for IoT Systems Cybersecurity, which has been funded by the UK EPSRC under grant number EP/S035362/1. This work has also benefited from the BIG ERAChair EU-funded project through Grant agreement ID: 952226. ITI/LARSyS is funded by FCT Project UIDB/50009/2020.

{\footnotesize \bibliographystyle{acm}
\bibliography{ref}}

\begin{thebibliography}{10}

\bibitem{HIPPA1}
Health insurance portability and accountability act of 1996 (hipaa), 2018.
\newblock https://www.cdc.gov/phlp/publications/topic/hipaa.html.

\bibitem{Maternity}
Discrimination during maternity leave and on return to work.
\newblock {\em Maternity Action\/} (2019).
\newblock https://maternityaction.org.uk/advice/discrimination-during-maternity-leave-and-on-return-to-work/.

\bibitem{International}
How menstruation apps are sharing your data.
\newblock {\em Privacy International\/} (2019).
\newblock https://privacyinternational.org/long-read/3196/no-bodys-business-mine-how-menstruations-apps-are-sharing-your-data.

\bibitem{ICObyDesign}
Data protection by design and default.
\newblock {\em ICO\/} (2020).
\newblock https://ico.org.uk/for-organisations/guide-to-data-protection/guide-to-the-general-data-protection-regulation-gdpr/accountability-and-governance/data-protection-by-design-and-default/.

\bibitem{WHO}
Sexual and reproductive health: Infertility.
\newblock {\em World Health Organization\/} (2020).
\newblock https://www.who.int/reproductivehealth/topics/infertility/keyissues/en/.

\bibitem{femtech1}
Femtech industry in interactive charts.
\newblock https://www.femtech.health/interactive-chartsh.

\bibitem{IranUN}
Iran death penalty threat for abortion unlawful: Un rights experts.
\newblock {\em United Nations\/} (2021).
\newblock https://news.un.org/en/story/2021/11/1105922.

\bibitem{Afnan}
{\sc Afnan, T., Zou, Y., Mustafa, M., Naseem, M., and Schaub, F.}
\newblock Aunties, strangers, and the {FBI:} online privacy concerns and experiences of muslim-american women.
\newblock In {\em Eighteenth Symposium on Usable Privacy and Security, {SOUPS} 2022, Boston, MA, USA, August 7-9, 2022\/} (2022), S.~Chiasson and A.~Kapadia, Eds., {USENIX} Association, pp.~387--406.

\bibitem{agrafiotis2018taxonomy}
{\sc Agrafiotis, I., Nurse, J.~R., Goldsmith, M., Creese, S., and Upton, D.}
\newblock A taxonomy of cyber-harms: Defining the impacts of cyber-attacks and understanding how they propagate.
\newblock {\em Journal of Cybersecurity 4}, 1 (2018), tyy006.

\bibitem{almeida2022}
{\sc Almeida, T., Shipp, L., Mehrnezhad, M., and Toreini, E.}
\newblock Bodies like yours: Enquiring data privacy in femtech.
\newblock In {\em NordiCHI Adjunct '22: Adjunct Proceedings of the 2022 Nordic Human-Computer Interaction Conference\/} (New York, NY, USA, 2022), Association for Computing Machinery.

\bibitem{minor}
{\sc Alvarez, P.}
\newblock House judiciary committee asks former orr director to clarify testimony on pregnant minors.
\newblock {\em CNN\/} (2019).
\newblock https://edition.cnn.com/2019/03/22/politics/scott-lloyd-pregnant-minors/index.html.

\bibitem{Bannon}
{\sc Bannon, M.~T.}
\newblock Why it is time to do away with the term femtech.

\bibitem{Jerry}
{\sc Beilinson, J.}
\newblock Glow pregnancy app exposed women to privacy threats.
\newblock Consumer Reports at consumerreports.org/mobile-security-software/glow-pregnancy-app-exposed-women-to-privacy-threats.

\bibitem{bergmann2020populism}
{\sc Bergmann, E.}
\newblock Populism and the politics of misinformation.
\newblock {\em Safundi 21}, 3 (2020), 251--265.

\bibitem{braun}
{\sc Braun, V., and Clarke, V.}
\newblock Thematic analysis.
\newblock In {\em APA handbook of research methods in psychology\/} (2012), vol.~2, American Psychological Association, p.~57–71.

\bibitem{brown2020supercharged}
{\sc Brown, E.}
\newblock Supercharged sexism: The triple threat of workplace monitoring for women.
\newblock {\em SSRN 3680861\/} (2020).

\bibitem{brown2021Femtech}
{\sc Brown, E.}
\newblock The femtech paradox: How workplace monitoring threatens women's equity.
\newblock {\em Jurimetrics\/} (2021).

\bibitem{Bussone}
{\sc Bussone, A., Kasadha, B., Stumpf, S., Durrant, A.~C., Tariq, S., Gibbs, J., Lloyd, K.~C., and Bird, J.}
\newblock Trust, identity, privacy, and security considerations for designing a peer data sharing platform between people living with hiv.
\newblock {\em Proc. ACM Hum.-Comput. Interact. 4}, CSCW2 (oct 2020).

\bibitem{cavoukian2009privacy}
{\sc Cavoukian, A., et~al.}
\newblock Privacy by design: The 7 foundational principles.
\newblock {\em Information and privacy commissioner of Ontario, Canada 5\/} (2009), 2009.

\bibitem{chan2021hidden}
{\sc Chan, S.}
\newblock Hidden but deadly: Stalkerware usage in intimate partner stalking.
\newblock {\em Introduction To Cyber Forensic Psychology: Understanding The Mind Of The Cyber Deviant Perpetrators\/} (2021), 45--66.

\bibitem{clarke2019editorial}
{\sc Clarke, V., Braun, V., Frith, H., and Moller, N.}
\newblock Editorial introduction to the special issue: Using story completion methods in qualitative research, 2019.

\bibitem{coopamootoo2022feel}
{\sc Coopamootoo, K., Mehrnezhad, M., and Toreini, E.}
\newblock " i feel invaded, annoyed, anxious and i may protect myself": Individuals' feelings about online tracking and their protective behaviour across gender and country.
\newblock {\em USENIX\/} (2022).

\bibitem{marketplace}
{\sc Cox, J.}
\newblock Data marketplace selling info about who uses period tracking apps, 2022.
\newblock {https://www.vice.com/en/article/v7d9zd/data-marketplace-selling-clue-period-tracking-data }.

\bibitem{crabtree2022privacy}
{\sc Crabtree, A., Haddadi, H., and Mortier, R.}
\newblock Privacy by design for the internet of things.
\newblock {\em Privacy by Design for the Internet of Things: Building Accountability and Security (2021). The Institution of Engineering and Technology: https://shop. theiet. org/privacy-by-design-for-the-internet-of-things\/} (2022).

\bibitem{crossley2005discrimination}
{\sc Crossley, M.}
\newblock Discrimination against the unhealthy in health insurance.
\newblock {\em U. Kan. L. Rev. 54\/} (2005), 73.

\bibitem{DeVito}
{\sc DeVito, M.~A., Walker, A.~M., and Birnholtz, J.}
\newblock 'too gay for facebook': Presenting lgbtq+ identity throughout the personal social media ecosystem.
\newblock {\em Proc. ACM Hum.-Comput. Interact. 2}, CSCW (nov 2018).

\bibitem{erickson2022you}
{\sc Erickson, J., Yuzon, J.~Y., and Bonaci, T.}
\newblock What you don’t expect when you’re expecting: Privacy analysis of femtech.
\newblock {\em IEEE Transactions on Technology and Society 3}, 2 (2022).

\bibitem{essen2022health}
{\sc Ess{\'e}n, A., Stern, A.~D., Haase, C.~B., Car, J., Greaves, F., Paparova, D., Vandeput, S., Wehrens, R., and Bates, D.~W.}
\newblock Health app policy: international comparison of nine countries’ approaches.
\newblock {\em NPJ digital medicine 5}, 1 (2022), 1--10.

\bibitem{Friedman}
{\sc Friedman, B., and Hendry, D.~G.}
\newblock {\em Value Sensitive Design: Shaping Technology with Moral Imagination}.
\newblock The MIT Press, 2019.

\bibitem{friedman2013value}
{\sc Friedman, B., Kahn, P.~H., Borning, A., and Huldtgren, A.}
\newblock Value sensitive design and information systems.
\newblock In {\em Early engagement and new technologies: Opening up the laboratory}. Springer, 2013, pp.~55--95.

\bibitem{Geeng}
{\sc Geeng, C., Harris, M., Redmiles, E., and Roesner, F.}
\newblock "like lesbians walking the perimeter": Experiences of {U.S}. {LGBTQ+} folks with online security, safety, and privacy advice.
\newblock In {\em 31st USENIX Security Symposium (USENIX Security 22)\/} (Boston, MA, Aug. 2022), USENIX Association, pp.~305--322.

\bibitem{Jessica}
{\sc Glenza, J.}
\newblock Revealed: women's fertility app is funded by anti-abortion campaigners.
\newblock {\em The Guardian\/} (2019).
\newblock theguardian.com/world/2019/may/30/revealed-womens-fertility-app-is-funded-by-anti-abortion-campaigners.

\bibitem{line}
{\sc Gro{\ss}, T.}
\newblock Why privacy is all but forgotten.
\newblock {\em Privacy Enhancing Technologies 2017}, 4 (2017), 97--118.

\bibitem{Drew}
{\sc Harwell, D.}
\newblock Is your pregnancy app sharing your intimate data with your boss?
\newblock {\em The Washington Post\/} (2019).
\newblock https://www.washingtonpost.com/technology/2019/04/10/tracking-your-pregnancy-an-app-may-be-more-public-than-you-think/.

\bibitem{Im}
{\sc Im, J., Schoenebeck, S., Iriarte, M., Grill, G., Wilkinson, D., Batool, A., Alharbi, R., Funwie, A., Gankhuu, T., Gilbert, E., and Naseem, M.}
\newblock Women's perspectives on harm and justice after online harassment.
\newblock {\em Proc. ACM Hum.-Comput. Interact. 6}, CSCW2 (nov 2022).

\bibitem{Joyce}
{\sc Joyce, J., and Sennett, P.}
\newblock 'femtech' – a market with a bright future and challenging terminology.

\bibitem{Karusala}
{\sc Karusala, N., Bhalla, A., and Kumar, N.}
\newblock Privacy, patriarchy, and participation on social media.
\newblock DIS '19, Association for Computing Machinery, p.~511–526.

\bibitem{kaye2014money}
{\sc Kaye, J.~J., McCuistion, M., Gulotta, R., and Shamma, D.~A.}
\newblock Money talks: tracking personal finances.
\newblock In {\em Proceedings of the SIGCHI Conference on Human Factors in Computing Systems\/} (2014), pp.~521--530.

\bibitem{collective}
{\sc Loi, M., and Christen, M.}
\newblock Two concepts of group privacy.
\newblock {\em Philosophy \& Technology\/} (2019), 1--18.

\bibitem{lupton1}
{\sc Lupton, D.}
\newblock ‘mastering your fertility’: The digitised reproductive citizen.
\newblock {\em Negotiating Digital Citizenship: Control, Contest and Culture\/} (2015).

\bibitem{McMillan}
{\sc McMillan, C.}
\newblock {Monitoring Female Fertility Through ‘Femtech’: The Need for a Whole-System Approach to Regulation}.
\newblock {\em Medical Law Review\/} (04 2022).
\newblock fwac006.

\bibitem{mcmillan2023rethinking}
{\sc McMillan, C.}
\newblock Rethinking the regulation of digital contraception under the medical devices regime.
\newblock {\em Medical Law International\/} (2023), 09685332231154581.

\bibitem{mehr}
{\sc Mehrnezhad, M., and Almeida, T.}
\newblock Caring for intimate data in fertility technologies.
\newblock In {\em Proceedings of the 2021 CHI Conference on Human Factors in Computing Systems\/} (New York, NY, USA, 2021), CHI '21, Association for Computing Machinery.

\bibitem{mehrnezhad2022can}
{\sc Mehrnezhad, M., Coopamootoo, K., and Toreini, E.}
\newblock How can and would people protect from online tracking?
\newblock {\em Proceedings on Privacy Enhancing Technologies 1\/} (2022), 105--125.

\bibitem{Mehr2022}
{\sc Mehrnezhad, M., Shipp, L., Almeida, T., and Toreini, E.}
\newblock Vision: Too little too late? do the risks of femtech already outweigh the benefits?
\newblock In {\em Proceedings of the 2022 European Symposium on Usable Security\/} (New York, NY, USA, 2022), EuroUSEC '22, Association for Computing Machinery, p.~145–150.

\bibitem{MindGap}
{\sc Mehrnezhad, M., van~der Merwe, T., and Catt, M.}
\newblock Mind the femtech gap: Regulation failings and exploitative systems.
\newblock {\em SOUPS Workshop (PEP)\/} (2023).

\bibitem{millett2001cookies}
{\sc Millett, L.~I., Friedman, B., and Felten, E.}
\newblock Cookies and web browser design: Toward realizing informed consent online.
\newblock In {\em Proceedings of the SIGCHI conference on Human factors in computing systems\/} (2001), pp.~46--52.

\bibitem{moniz2023intimate}
{\sc Moniz, D.~P., Mehrnezhad, M., and Almeida, T.}
\newblock Intimate data: Exploring perceptions of privacy and privacy-seeking behaviors through the story completion method.
\newblock In {\em Human-Computer Interaction -- INTERACT 2023\/} (Cham, 2023), J.~Abdelnour~Nocera, M.~Krist{\'i}n~L{\'a}rusd{\'o}ttir, H.~Petrie, A.~Piccinno, and M.~Winckler, Eds., Springer Nature Switzerland, pp.~533--543.

\bibitem{nicol2022revealing}
{\sc Nicol, E., Briggs, J., Moncur, W., Htait, A., Carey, D.~P., Azzopardi, L., and Schafer, B.}
\newblock Revealing cumulative risks in online personal information: a data narrative study.
\newblock {\em Proceedings of the ACM on Human-Computer Interaction 6}, CSCW2 (2022), 1--25.

\bibitem{AntiAbortion}
{\sc Page, C.}
\newblock As roe v. wade reversal looms, should you delete your period-tracking app?, 2022.
\newblock https://techcrunch.com/2022/05/05/roe-wade-privacy-period-tracking.

\bibitem{pennycook2021shifting}
{\sc Pennycook, G., and et~al.}
\newblock Shifting attention to accuracy can reduce misinformation online.
\newblock {\em Nature\/} (2021).

\bibitem{peppet2011unraveling}
{\sc Peppet, S.~R.}
\newblock Unraveling privacy: The personal prospectus and the threat of a full-disclosure future.
\newblock {\em Nw. UL Rev.\/} (2011).

\bibitem{powles2017google}
{\sc Powles, J., and Hodson, H.}
\newblock Google deepmind and healthcare in an age of algorithms.
\newblock {\em Health and technology\/} (2017).

\bibitem{rosas}
{\sc Rosas, C.}
\newblock The future is femtech: Privacy and data security issues surrounding femtech applications.
\newblock {\em Hastings Business Law Journal 15}, 2 (2019).

\bibitem{rosenbaum2009insurance}
{\sc Rosenbaum, S.}
\newblock Insurance discrimination on the basis of health status: An overview of discrimination practices, federal law, and federal reform options: Executive summary.
\newblock {\em Journal of Law, Medicine \& Ethics 37}, S2 (2009), 101--120.

\bibitem{ryan2009device}
{\sc Ryan, W., Stolterman, E., Jung, H., Siegel, M., Thompson, T., and Hazlewood, W.~R.}
\newblock Device ecology mapper: a tool for studying users' ecosystems of interactive artifacts.
\newblock In {\em CHI'09 Extended Abstracts on Human Factors in Computing Systems}. 2009, pp.~4327--4332.

\bibitem{Sambasivan}
{\sc Sambasivan, N., Batool, A., Ahmed, N., Matthews, T., Thomas, K., Gayt\'{a}n-Lugo, L.~S., Nemer, D., Bursztein, E., Churchill, E., and Consolvo, S.}
\newblock "they don't leave us alone anywhere we go": Gender and digital abuse in south asia.
\newblock In {\em Proceedings of the 2019 CHI Conference on Human Factors in Computing Systems\/} (New York, NY, USA, 2019), CHI '19, Association for Computing Machinery, p.~1–14.

\bibitem{Sambasivan2}
{\sc Sambasivan, N., Checkley, G., Ahmed, N., and Batool, A.}
\newblock Gender equity in technologies: Considerations for design in the global south.
\newblock {\em Interactions 25}, 1 (dec 2017), 58–61.

\bibitem{Sambasivan1}
{\sc Sambasivan, N., Checkley, G., Batool, A., Ahmed, N., Nemer, D., Gayt\'{a}n-Lugo, L.~S., Matthews, T., Consolvo, S., and Churchil, E.}
\newblock "privacy is not for me, it's for those rich women": Performative privacy practices on mobile phones by women in south asia.
\newblock In {\em Proceedings of the Fourteenth USENIX Conference on Usable Privacy and Security\/} (USA, 2018), SOUPS '18, USENIX Association, p.~127–142.

\bibitem{scatterday2022no}
{\sc Scatterday, A.}
\newblock This is no ovary-action: Femtech apps need stronger regulations to protect data and advance public health goals.
\newblock {\em North Carolina Journal of Law \& Technology 23}, 3 (2022), 636.

\bibitem{shipp2020private}
{\sc Shipp, L., and Blasco, J.}
\newblock How private is your period?: A systematic analysis of menstrual app privacy policies.
\newblock {\em Proceedings on Privacy Enhancing Technologies 4\/} (2020), 491--510.

\bibitem{Shoichet}
{\sc Shoichet, C.}
\newblock In a horrifying history of forced sterilizations, some fear the us is beginning a new chapter, 2020.
\newblock https://edition.cnn.com/2020/09/16/us/ice-hysterectomy-forced-sterilization-history/index.html.

\bibitem{Slupska:2019:1746-451X:83}
{\sc Slupska, J.}
\newblock Safe at home: Towards a feminist critique of cybersecurity.
\newblock {\em St Antony's International Review 15}, 1 (2019), 83--100.

\bibitem{slupska2022they}
{\sc Slupska, J., Cho, S., Begonia, M., Abu-Salma, R., Prakash, N., and Balakrishnan, M.}
\newblock " they look at vulnerability and use that to abuse you'': Participatory threat modelling with migrant domestic workers.
\newblock In {\em 31st USENIX Security Symposium (USENIX Security 22)\/} (2022), pp.~323--340.

\bibitem{slupska2021participatory}
{\sc Slupska, J., Dawson~Duckworth, S.~D., Ma, L., and Neff, G.}
\newblock Participatory threat modelling: Exploring paths to reconfigure cybersecurity.
\newblock In {\em Extended Abstracts of the 2021 CHI Conference on Human Factors in Computing Systems\/} (2021), pp.~1--6.

\bibitem{slupska2021reconfigure}
{\sc Slupska, J., Duckworth, S.~D., and Neff, G.}
\newblock Reconfigure: feminist action research in cybersecurity.

\bibitem{stevens2021cyber}
{\sc Stevens, F., Nurse, J.~R., and Arief, B.}
\newblock Cyber stalking, cyber harassment, and adult mental health: A systematic review.
\newblock {\em Cyberpsychology, Behavior, and Social Networking 24}, 6 (2021), 367--376.

\bibitem{strohmayer2022safety}
{\sc Strohmayer, A., Bellini, R., and Slupska, J.}
\newblock Safety as a grand challenge in pervasive computing: Using feminist epistemologies to shift the paradigm from security to safety.
\newblock {\em IEEE Pervasive Computing\/} (2022).

\bibitem{Troiano}
{\sc Troiano, G.~M., Wood, M., and Harteveld, C.}
\newblock "and this, kids, is how i met your mother": Consumerist, mundane, and uncanny futures with sex robots.
\newblock In {\em Proceedings of the 2020 CHI Conference on Human Factors in Computing Systems\/} (New York, NY, USA, 2020), CHI '20, Association for Computing Machinery, p.~1–17.

\bibitem{Tseng}
{\sc Tseng, E., Sabet, M., Bellini, R., Sodhi, H.~K., Ristenpart, T., and Dell, N.}
\newblock Care infrastructures for digital security in intimate partner violence.
\newblock In {\em Proceedings of the 2022 CHI Conference on Human Factors in Computing Systems\/} (New York, NY, USA, 2022), CHI '22, Association for Computing Machinery.

\bibitem{valente2019stealing}
{\sc Valente, J., Wynn, M.~A., and Cardenas, A.~A.}
\newblock Stealing, spying, and abusing: Consequences of attacks on internet of things devices.
\newblock {\em IEEE Security \& Privacy 17}, 5 (2019), 10--21.

\bibitem{kerry}
{\sc van~der Berch, K.}
\newblock Courts’ struggle with infertility: the impact of hall v. nalco on infertility-related employment discrimination.
\newblock {\em University of Colorado Law Review 81}, 2 (2010).

\bibitem{vertesi2016data}
{\sc Vertesi, J., Kaye, J., Jarosewski, S.~N., Khovanskaya, V.~D., and Song, J.}
\newblock Data narratives: Uncovering tensions in personal data management.
\newblock In {\em Proceedings of the 19th ACM conference on Computer-Supported cooperative work \& social computing\/} (2016), pp.~478--490.

\bibitem{Carissa}
{\sc Véliz, C.}
\newblock Privacy is power: Why and how you should take back control of your data.
\newblock Bantam Press.

\bibitem{wambsganss2021ethical}
{\sc Wambsganss, T., H{\"o}ch, A., Zierau, N., and S{\"o}llner, M.}
\newblock Ethical design of conversational agents: towards principles for a value-sensitive design.
\newblock In {\em International Conference on Wirtschaftsinformatik\/} (2021), Springer, pp.~539--557.

\bibitem{Ash}
{\sc Watson, A., and Lupton, D.}
\newblock What happens next? using the story completion method to surface the affects and materialities of digital privacy dilemmas.
\newblock {\em Sociological Research Online 1}, 7 (2022), 13607804221084343.

\bibitem{Wood}
{\sc Wood, M., Wood, G., and Balaam, M.}
\newblock "they're just tixel pits, man": Disputing the 'reality' of virtual reality pornography through the story completion method.
\newblock In {\em Proceedings of the 2017 CHI Conference on Human Factors in Computing Systems\/} (New York, NY, USA, 2017), CHI '17, Association for Computing Machinery, p.~5439–5451.

\bibitem{xu2012value}
{\sc Xu, H., Crossler, R.~E., and B{\'e}Langer, F.}
\newblock A value sensitive design investigation of privacy enhancing tools in web browsers.
\newblock {\em Decision support systems 54}, 1 (2012), 424--433.

\bibitem{xu2008alleviating}
{\sc Xu, H., Irani, N., Zhu, S., and Xu, W.}
\newblock Alleviating parental concerns for children's online privacy: a value sensitive design investigation.

\end{thebibliography}
\appendix
\section{User Study Template for Online Survey}
[General description]

[1-3: Demo questions]

4. In your own words, please describe what are FemTech and what do they do? 

5. What forms of FemTech do you use (or have used)? Please give examples and how and when do (did) you use them?

6. How did you learn about these technologies? (multiple choices)
\begin{itemize}
\item Personal research
\item Online ad
\item Traditional ad (leaflet, ...)
\item Family/Friend
\item Colleague 
\item School/University
\item NGO/Charity
\item Health centre/clinic or NHS
\item TV/Radio
\item Others (please name)
\end{itemize}

7. What type of data do you enter in these technologies about yourself (e.g. name, email, medical records, sexual activities, etc.)? Please list all. 

8. What type of data do you enter in these technologies about yourself? (multiple choices)
\begin{itemize}
\item Name
\item Photo
\item Age/Gender
\item Contact info (mobile, email, ads)
\item Social media (Facebook, Instagram, Twitter, ...)
\item Lifestyle (weight, height, diet, exercise, sleep, ...) 
\item Period info (start, end, cycle days, evaluation days, ...)
\item Sexual activities (type, date, contraceptive, orgasm, ...)
\item Reproductive organs (cervical mucus, ... )
\item Pregnancy info (test result, medical info, weight gain, body change, ...)
\item Baby info (nursing, sleep cycles, ...)
\item Physical symptoms (headache, insomnia, constipation, ...)
\item Emotional symptoms (happy, upset, energised, tired, anxious, ...)
\item Medical info (medication type, dose, blood pressure, ...)
\item Others (please name)
\end{itemize}

9.	Why do you use such technologies? Please explain.

10. In your opinion, what are the benefits of using such technologies? Please explain. 

11. In your opinion, what are the benefits of using such technologies? (multiple choices)
\begin{itemize}
    \item They are cost-effective
    \item Everyone can have access to them
    \item They are accurate
    \item It is easy to use them
    \item They can be personalised
    \item They are non-intrusive to body
    \item They increase my knowledge about my body
     \item Others (please specify)
\end{itemize}

12. In your opinion, what are the disadvantages of using such technologies? Please explain. 

13.	In your opinion, what are the disadvantages of using such technologies? (multiple choices)
\begin{itemize}
    \item They are expensive
    \item They are not accessible to all user groups
    \item Using such technologies are not easy for certain users 
    \item They are not accurate 
    \item They only work for certain user groups and have bias
    \item They share user information with third parties or sell to them 
     \item Others (please specify)
\end{itemize}

14. Have you (or someone you know) ever experienced an unpleasant incident as a result of using such technologies? (e.g. unwanted pregnancy or stress of losing fertile days). Please explain.

15.	Who do you think has access to your data when using FemTech? (multiple choices)
\begin{itemize}
    \item Your GP/NHS
    \item Your partner
    \item Your family/friend/co-worker
    \item Your employer
    \item The company making the app/device
    \item Operating system of mobile/PC (e.g. Apple, Google. Microsoft)
    \item Third parties such as medical clinics and insurance companies 
\item Third parties such as advertising companies
\item The government
\item The public
     \item Others (please name)
\end{itemize}

16. Have you (or someone you know) ever experienced a cybersecurity/privacy incident as a result of using such technologies? (e.g. sharing your data without consent). Please explain. 

17. What do you generally do to protect your online privacy and security? 
Please explain.

18.	What to you specifically do to protect your FemTech privacy and security? Please explain.

19.	In your opinion, what sorts of security and privacy features should be included in such technologies to make you use them without risk and fear? Please explain. 

20. If you have any other comments about FemTech, please write them here. 
 

\section{Story Completion Activity}
\textbf{Description}: Welcome and thanks for taking part in this part of our study!
In this activity, you will find three hypothetical scenarios. Each one will introduce you to a topic related to using FemTech. We invite you to write a story in response to each scenario.

Please spend around 15 minutes writing on each scenario (a total of 45 minutes exercise).  

\textbf{Section 1 of 3:}
Alex downloads a free period tracking app to keep an eye on their fertility window. The next day Alex receives multiple email ads about baby’s clothing and toys.

What does Alex think? What happens next?

\textbf{Section 2 of 3:}
Morgan goes online and browses the internet, searching for common symptoms of menopause. The next time Morgan logs on to Instagram and Facebook, both feeds are full of ads for boosting sex drive and weight loss.

What does Morgan think? What happens next?

\textbf{Section 3 of 3:}
Ashley has been using a ‘smart’ breast pump that connects to an app and is shared within the family. This is a breast pump that can track breastfeeding and give tips to support parents in meeting their goals. In the meantime, Ashley’s partner has received an alert regarding the baby’s feeding schedule and weight and feels upset that it is not according to their plans.

What does Ashley think? What happens next?

\textbf{Next steps:} We will be in touch for the next part of the study (which will be an online one-to-one interview as instructed in the email). If you have any comments or concerns, please communicate with us as instructed in the email. 
\section{Examples of Participants' Drawings}
\begin{figure}[h]
\includegraphics[scale=.35]{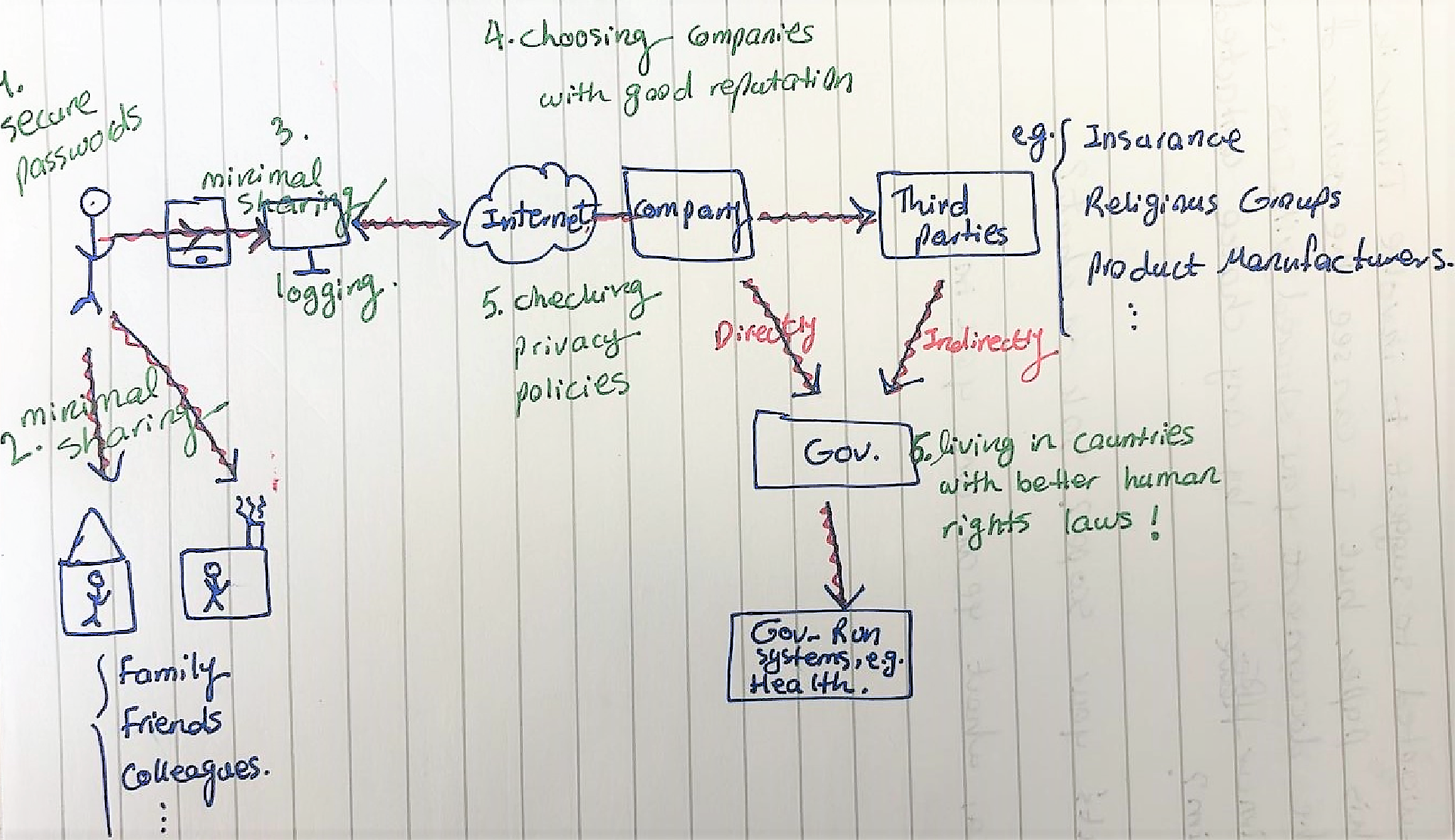}
\includegraphics[scale=.6]{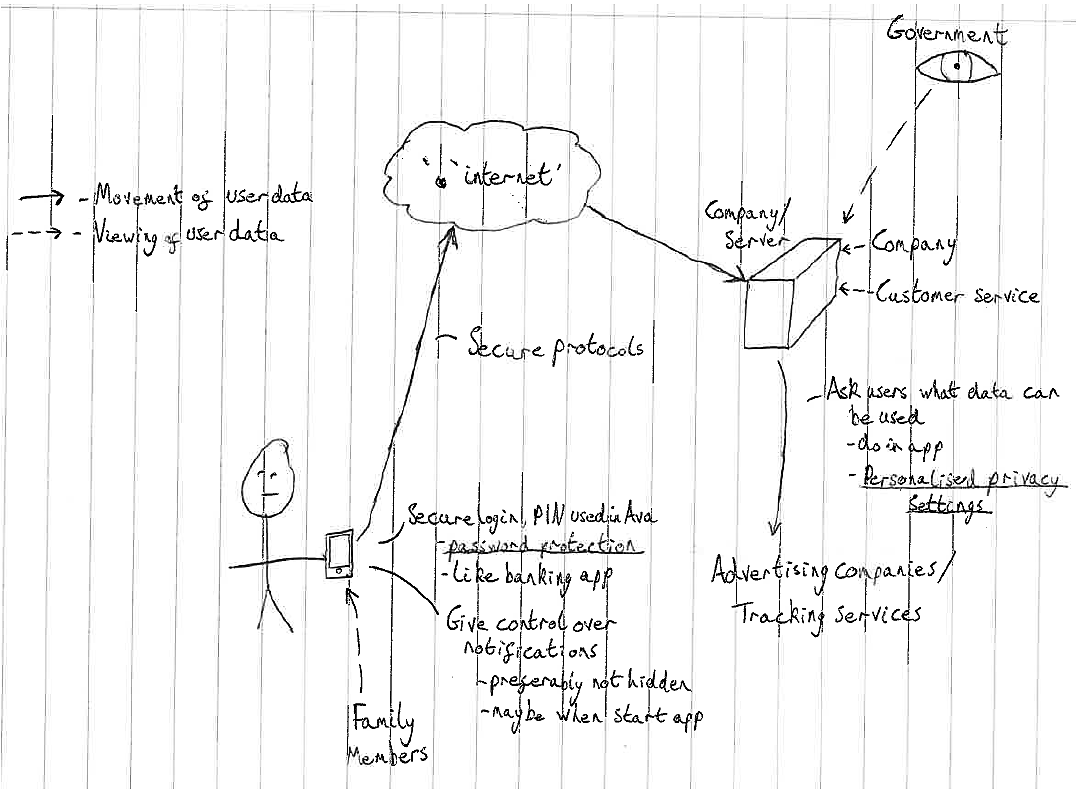}
\caption{Examples of drawings of FemTech systems by participants in the Interviews}
\label{Eco}
\end{figure}
\end{document}